\documentclass[aps,prb,twocolumn,superscriptaddress,showpacs]{revtex4-1}
\usepackage{hyperref}
\usepackage{graphicx}
\usepackage{revtex4-1}
\usepackage{units}

\mathchardef\mhyphen="2D

\newcommand{\FT}{Fe$_{1+y}$Te}
\newcommand{\FTS}{Fe$_{1+y}$Te$_{1-x}$Se$_{x}$}
\newcommand{\samp}{Fe$_{1.08}$Te}

\newcommand{\BFAN}{BaFe$_{2-x}$Ni$_{x}$As$_{2}$}

\newcommand{\kf}{k$_f$}		
\newcommand{\Ef}{E$_f$}		
\newcommand{\TN}{T$_N$}		
\newcommand{\Tc}{T$_c$}		
\newcommand{\Qvec}{$\mathbf{q}$}
\newcommand{\Qafmft}{$\mathbf{q}_{AFM}$ = [$\nicefrac{1}{2}$ , 0, $\nicefrac{1}{2}$]}
\newcommand{\Qafm}{$\mathbf{q}_{AFM}$}
\newcommand{\Qinc}{$\mathbf{q}_{inc}$}	
\newcommand{\Qorder}{$\mathbf{q}_{ordering}$}	
\newcommand{\sampTN}{T$_N$ = 67.5 K}	

\begin{document}

\title{Competition between commensurate and incommensurate magnetic ordering in \FT}



\author{D.~Parshall}
	\email{dan.parshall@colorado.edu}   
\author{G.~Chen}
	\affiliation{University of Colorado, Department of Physics, Boulder, CO 80309}
\author{L.~Pintschovius}
	\affiliation{Karlsruhe Institut f\"ur Technologie, Institut f\"ur Festk\"orperphysik, P.O.B. 3640, D-76021 Karlsruhe, Germany}
\author{D.~Lamago}
	\affiliation{Karlsruhe Institut f\"ur Technologie, Institut f\"ur Festk\"orperphysik, P.O.B. 3640, D-76021 Karlsruhe, Germany}
	\affiliation{Laboratoire L\'eon Brillouin, CEA-Saclay, F-91191 Gif-sur-Yvette Cedex, France}
\author{Th.~Wolf}
	\affiliation{Karlsruhe Institut f\"ur Technologie, Institut f\"ur Festk\"orperphysik, P.O.B. 3640, D-76021 Karlsruhe, Germany}
\author{L.~Radzihovsky}
	\affiliation{University of Colorado, Department of Physics, Boulder, CO 80309}
\author{D.~Reznik}
	\affiliation{University of Colorado, Department of Physics, Boulder, CO 80309}

\date{\today}

\begin{abstract}
The \FTS\ compounds belong to the family of iron-based high temperature superconductors, in which superconductivity often appears upon doping antiferromagnetic parent compounds.  Unlike other Fe-based superconductors (in which the antiferromagnetic order is at the Fermi surface nesting wavevector [\nicefrac{1}{2},\nicefrac{1}{2},1]), \FT\ orders at a different wavevector, [\nicefrac{1}{2}, 0, \nicefrac{1}{2}].  Furthermore, the ordering wavevector depends on $y$, the occupation of interstitial sites with excess iron; the origin of this behavior is controversial.  Using inelastic neutron scattering on \samp, we find incommensurate magnetic fluctuations above the N\'eel temperature, even though the ordered state is bicollinear and commensurate with gapped spin waves.  This behavior can be understood in terms of a competition between commensurate and incommensurate order, which we explain as a lock-in transition caused by the magnetic anisotropy.
\end{abstract}


\maketitle


\begin{figure}[br]
   \centering
   \includegraphics[scale=0.5]{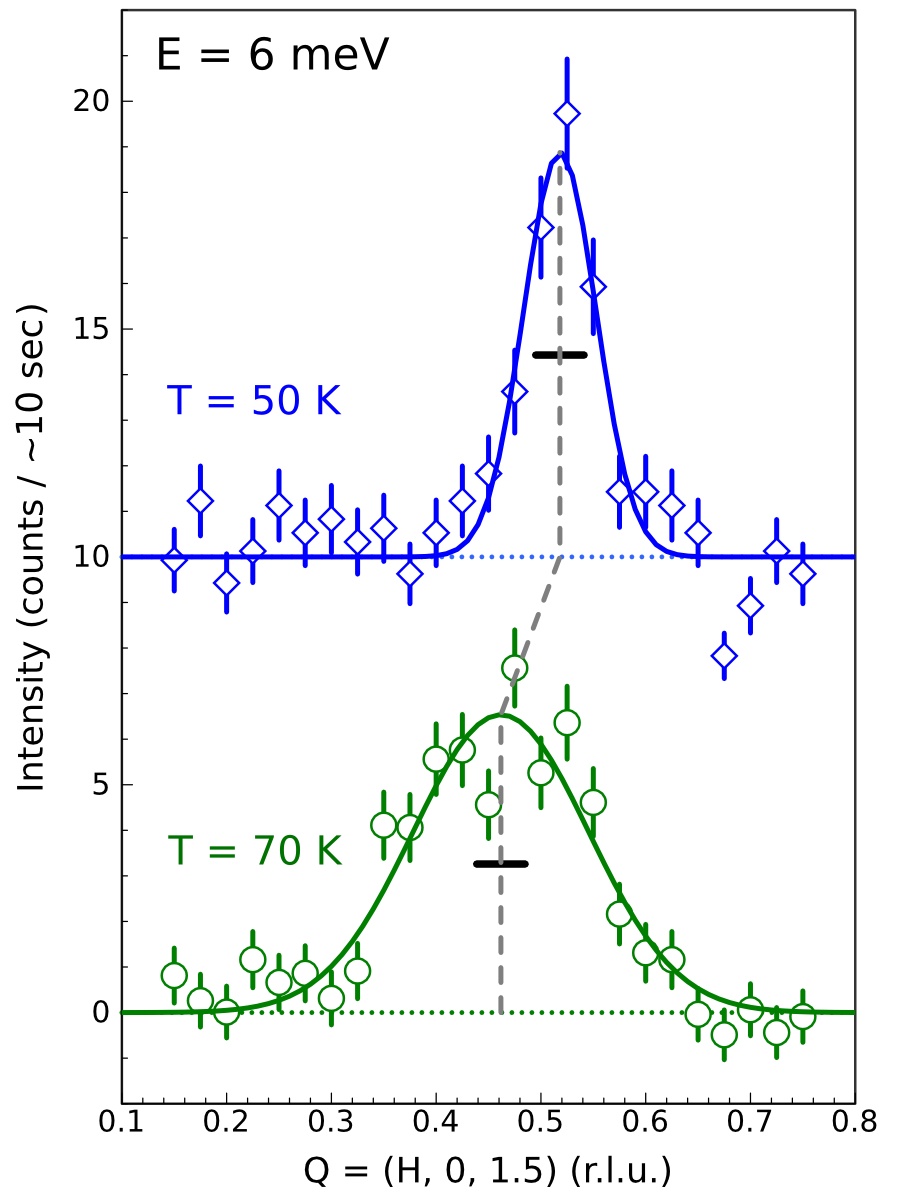} 
    \caption{(color online) Scans through [H, 0, \nicefrac{1}{2}] with E = 6 meV, taken above and below \sampTN.  The black horizontal bar shows the estimated resolution width.  The shift of the spin fluctuations from the commensurate to incommensurate position (when warming from the ordered to paramagnetic state) has not been previously reported.}
   \label{fig:H-scan}
\end{figure}

Superconductivity in the recently-discovered iron-based superconductors (FeSCs) \cite{Kamihara2008} often appears at high transition temperatures.  These compounds contain a simple square lattice of iron atoms, coordinated with pnictogen or chalcogen atoms forming planes of tetrahedra. Interplanar layers differ between families, consisting of either metal oxides, alkaline earth atoms, alkali atoms, or nothing at all as in the ``11'' compounds on which we focus here.  The parent compounds of most families become orthorhombic and antiferromagnetic (AFM) at low temperatures, and superconductivity can be induced by doping with electrons, holes, isoelectronically, or by the application of pressure (see, e.g., \cite{Paglione2010}).  Their high-\Tc\ superconductivity may be related to the magnetic order \cite{Mazin08}, which seems to be the result of Fermi surface nesting \cite{Singh08}.  There is also intrinsic interest in magnetism in these materials, but it has not been as extensively explored.

While nesting explains many experiments, the magnetic order in the $x = 0$ endpoint of the \FTS\ series is a highly unusual commensurate ``bicollinear'' magnetic structure \cite{Bao09} with a wavevector along \Qafmft.  Neither density functional theory (DFT) \cite{Subedi08} nor photoemission measurements \cite{Xia2009} find any evidence for nesting at this wavevector, which  indicates the presence of local moments.  These moments must interact with each other via competing exchange interactions \cite{Turner09}, which should lead to incommensurate order. The mechanism behind the commensurate bicollinear order is one of the main unresolved issues in the effort to understand the FeSCs.

Here we report results of a detailed inelastic neutron scattering (INS) investigation of the region near \Qafm\ in a high-quality single-crystal sample of \samp.  We find strong evidence for competition between commensurate and incommensurate ordering in the form of spin excitations which abruptly shift from the incommensurate \Qinc\ $\approx$ [0.45, 0, 0.5] to the commensurate wavevector \Qafmft\ when passing below the N\'eel temperature \TN\ (see Fig. \ref{fig:H-scan}).  We interpret this unusual behavior in terms of a lock-in transition driven by crystalline anisotropy, which causes bicollinear order.

\begin{figure}[h]
   \centering
   \includegraphics[scale=0.45]{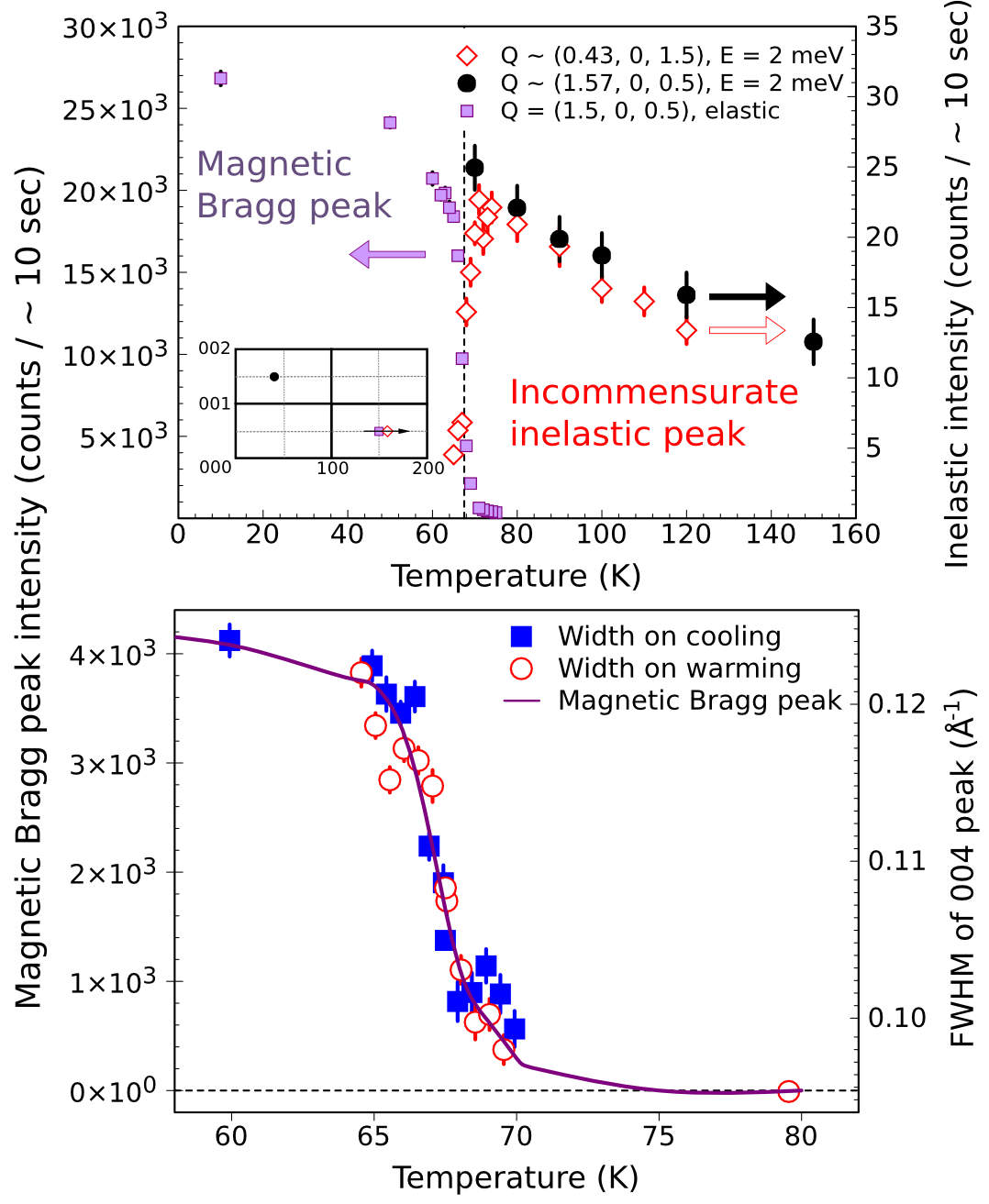} 
    \caption{(color online) Temperature dependence of the magnetic and structural behavior.  
{\bf (a)} Peak intensities of the incommensurate magnetic excitation and the [1.5, 0, 0.5] magnetic Bragg peak as a function of temperature.  Inset: map of reciprocal space, showing the locations of the scans.  
{\bf (b)} Width of the [0, 0, 4] nuclear Bragg peak (an indicator of monoclinic splitting) as a function of temperature, showing the structural transition at 67.5 K.  This is plotted together with the intensity of the [2.5, 0, 0.5] magnetic Bragg peak, demonstrating the close coincidence between the structural and magnetic transitions.}
   \label{fig:mag_vs_temp}
\end{figure}

The INS measurements were performed on a single crystal of \samp\ ($\approx$ 0.1 cc, $\approx$ 2 g, mosaic spread 2$^{\circ}$).  It was grown by the Bridgman technique, and the excess iron content was determined by Patterson refinement of single-crystal x-ray diffraction data to be $y$ = 0.08.  The N\'eel temperature was defined as the steepest slope of the magnetic Bragg peak intensity with temperature, was found to be \sampTN, consistent with other reports \cite{Liu10, Lipscombe11, StockRodriguez} for this value of $y$.  The crystal structure at room temperature is tetragonal (space group 129, P4/nmm) with lattice constants $a$ = $b$ = 3.823(3) \AA\  and $c$ = 6.282(6) \AA.  Although this compound becomes monoclinic (space group 11, P2$_1$/m) at low temperature \cite{Li09-1}, the primary effect is a shift of the Te atoms within the unit cell \cite{Bao09}; the rotation of the $c$-axis away from $90^\circ$ is quite small, and amounts to a broadening of certain Bragg peaks (see Fig. \ref{fig:mag_vs_temp}b), so we describe the measurements in the tetragonal [HHL] notation.

The neutron measurements were performed on the 1T1 triple-axis spectrometer at the Laboratoire L\'eon Brillouin, Saclay, France.  The sample was mounted in a standard displex cryostat (base temperature T = 11 K) in the [H0L] plane. The measurements were done with vertically and horizontally focusing PG crystals as monochromator and analyzer, respectively.  A graphite filter was used to reduce $\lambda$/2 contamination.  The scans were performed using a fixed final wavevector of \kf\ = 2.662 \AA\ (\Ef\ = 14.7 meV), corresponding to an energy resolution of 0.8 meV at the elastic line.  Fitting was done using the Fityk program \cite{Wojdyr10}.  All constant-\Qvec\ scans were fit with a background function (either constant or linear) and Gaussians for the peaks.  Elastic scans were fit with a constant background and Voigt profile.

We measured inelastic and elastic magnetic scattering from our sample near  \Qafm, obtaining a detailed picture of the evolution of the magnetic order, the crystal structure, and the magnetic excitations as a function of temperature in the vicinity of the first-order phase transition from the high-temperature paramagnetic to the low-temperature AFM phase. 

Figure \ref{fig:H-scan} shows constant-energy scans, at E = 6 meV, along [H, 0, \nicefrac{1}{2}] above and below \TN. At 50 K the peak is commensurate and comparable in width to the resolution function, whereas at 70 K it is broad and incommensurate. 

Figure \ref{fig:mag_vs_temp}a details the peak scattering intensity at 2 meV, which is well below the low-T spin gap, taken along the [H, 0, \nicefrac{1}{2}]-direction.  This is plotted together with the temperature dependence of the magnetic Bragg peak. The phase transition at 
67.5 K 
is clearly evident: the intensity at 2 meV builds up on cooling towards the phase transition, then drops abruptly as the spin gap opens and the magnetic Bragg peak appears.  The rapid increase of magnetic intensity is consistent with a first-order phase transition.  We also measured the temperature dependence of the mosaic of the [004] lattice reflection, which is proportional to the amount of the monoclinic distortion.  Figure \ref{fig:mag_vs_temp}b demonstrates that the magnetic transition is concurrent with the structural phase transition to the monoclinic phase.

\begin{figure}[t]
   \centering
   \includegraphics[scale=0.45]{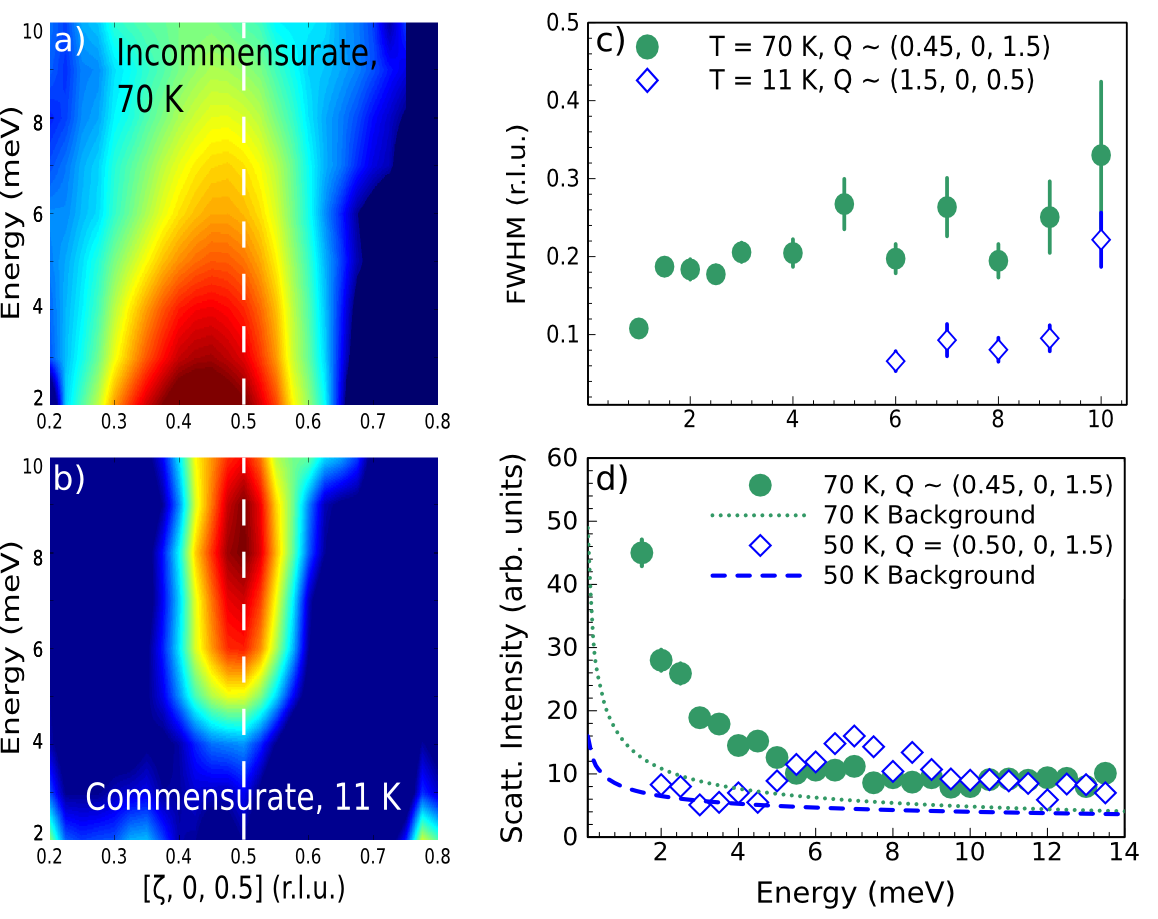} 
    \caption{ (color online) Energy dependence of the spin excitations. 
{\bf (a,b)} $S(q,\omega)$ at T = 70 K (a) and 11 K (b).  The data have been smoothed and background-subtracted.  Intensity is plotted on a logarithmic scale (arb. units).
{\bf (c)} Linewidths of the incommensurate (solid green circles) and commensurate (open blue diamonds) peaks, as a function of energy.
{\bf (d)} Constant-Q scans taken at the center of the incommensurate and commensurate spin excitations.}
   \label{fig:energy_dep}
\end{figure}

Figure \ref{fig:energy_dep} shows $S(q,\omega)$, and summarizes the energy- and wavevector-dependence of the magnetic inelastic scattering above and below \TN. Above \TN\ it is incommensurate, broad, and ungapped. Below \TN\ it is commensurate, narrow, and gapped.  Both have nearly vertical dispersion within the experimental uncertainty.  

\begin{figure}[b]
   \centering
   \includegraphics[scale=0.65]{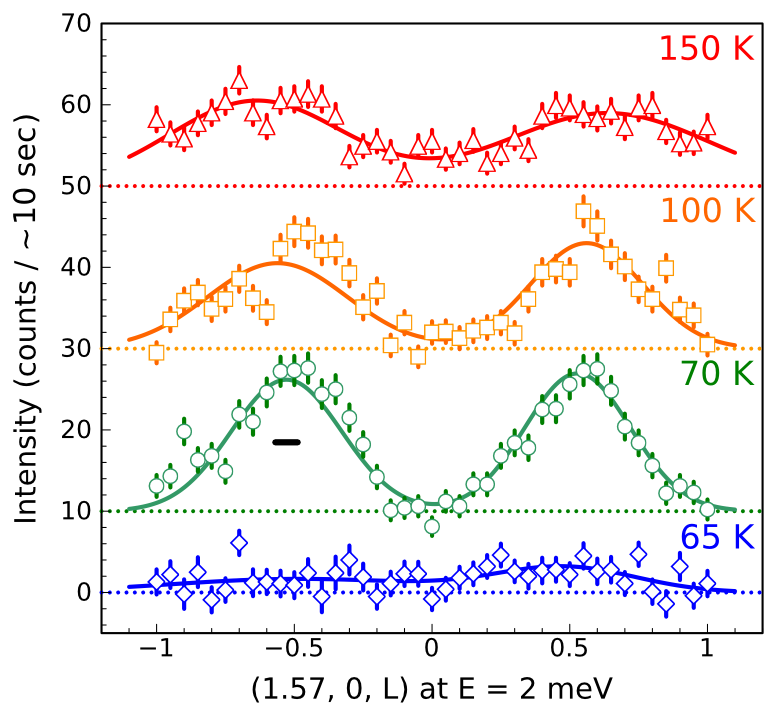} 
    \caption{ (color online) Scans along the $c$-axis direction.  The black horizontal bar shows the estimated FWHM of the resolution function.  The integrated intensity is approximately constant for temperatures above \TN, indicating that the number of spins is constant, but the correlation length decreases with T.}
   \label{fig:L-scan}
\end{figure}

The temperature dependence of the signal with L is shown in Fig. \ref{fig:L-scan}.  The magnetic scattering at 2 meV is nearly absent at 65 K, because of the spin gap opening at the phase transition.  Upon heating in the paramagnetic phase, the signal broadens along the L-direction, although the integrated intensity stays roughly the same.  The signal broadens only slightly in the H-direction (not shown).  Based on this observation, we conclude that the buildup of intensity in H-integrated scans at 2 meV towards \TN\ shown in Fig. \ref{fig:mag_vs_temp} results from an increase in the magnetic correlation length along the $c$-axis, as opposed to an increase of total scattering.

The most important feature of the observed behavior is that incommensurate fluctuations give rise to commensurate order, or more generally, fluctuations appear at one wavevector while ordering occurs at another.
Our measurements demonstrate that, somewhat surprisingly, the shift in the inelastic peak is energy independent all the way up to the highest observed energy of 10 meV.
Existing theories do not provide a cromulent explanation for the observed behavior.  According to the orbital scheme as discussed, for example, in Ref. \cite{Turner09}, in the tetragonal phase the highest energy electron can occupy either d$_{xz}$ or d$_{yz}$, orbitals which are otherwise empty.  This orbital degeneracy implies that the entire spin system is free to rotate in the $x$-$y$ plane with no in-plane anisotropy.  In such a model, the bicollinear spin order results from a combination of anisotropic exchange and the biquadratic spin interaction.  However, the predicted gapless spin-wave dispersion is inconsistent with the large spin gap that we and others have observed \cite{Lipscombe11, StockRodriguez}.  Using DFT, Ma \emph{et al}. \cite{Ma09} have found the lowest-energy \emph{commensurate} phase to be bicollinear, even when the lattice is tetragonal.  They did not calculate whether incommensurate ordering is more favorable than commensurate (DFT requires calculations using large supercells to understand the competition between commensurate and incommensurate order).

Our results (as well as those of previous experiments) may be captured by a simple model \cite{Choi_Chen_Radzihovsky}, in which the spin rotation symmetry is explicitly broken by the single-ion anisotropy.  The low-temperature monoclinic phase features both an orthorhombic distortion with shortening along the $b$-axis, as well as a monoclinic shearing of the Te-plans along the $a$-axis.  Together these distortions break the symmetry along the $a$- and $b$-axes.  Thus the crystal field environment becomes anisotropic, and the degeneracy between the d$_{xz}$ and d$_{yz}$ orbitals is lifted.  

This single-ion anisotropy is at the heart of magnetoelastic coupling, 
causing an Ising-type behavior in which the spins are locked along the $b$-axis, which in turn opens a gap in the spin-wave spectrum (in agreement with experiments).  In the tetragonal or orthorhombic phase with incommensurate magnetic order, the exchange interaction becomes dominant, and the magnetoelastic coupling does not lift the degeneracy between the d$_{xz}$ and d$_{yz}$ orbitals.  In such a Heisenberg spin system, the various Js set the periodicity of the spin system, but the spins are free to rotate with respect to the lattice, and thus the spin excitations are gapless, and the fluctuating moments should be isotropic.  However, spin fluctuations in the related system \BFAN\ were found to have a preferred axis \cite{Lipscombe10}, lending support to our interpretation of a Ising-type behavior in which the spin axis is coupled to the lattice.

\begin{figure}[b]
   \centering
   \includegraphics[scale=0.35]{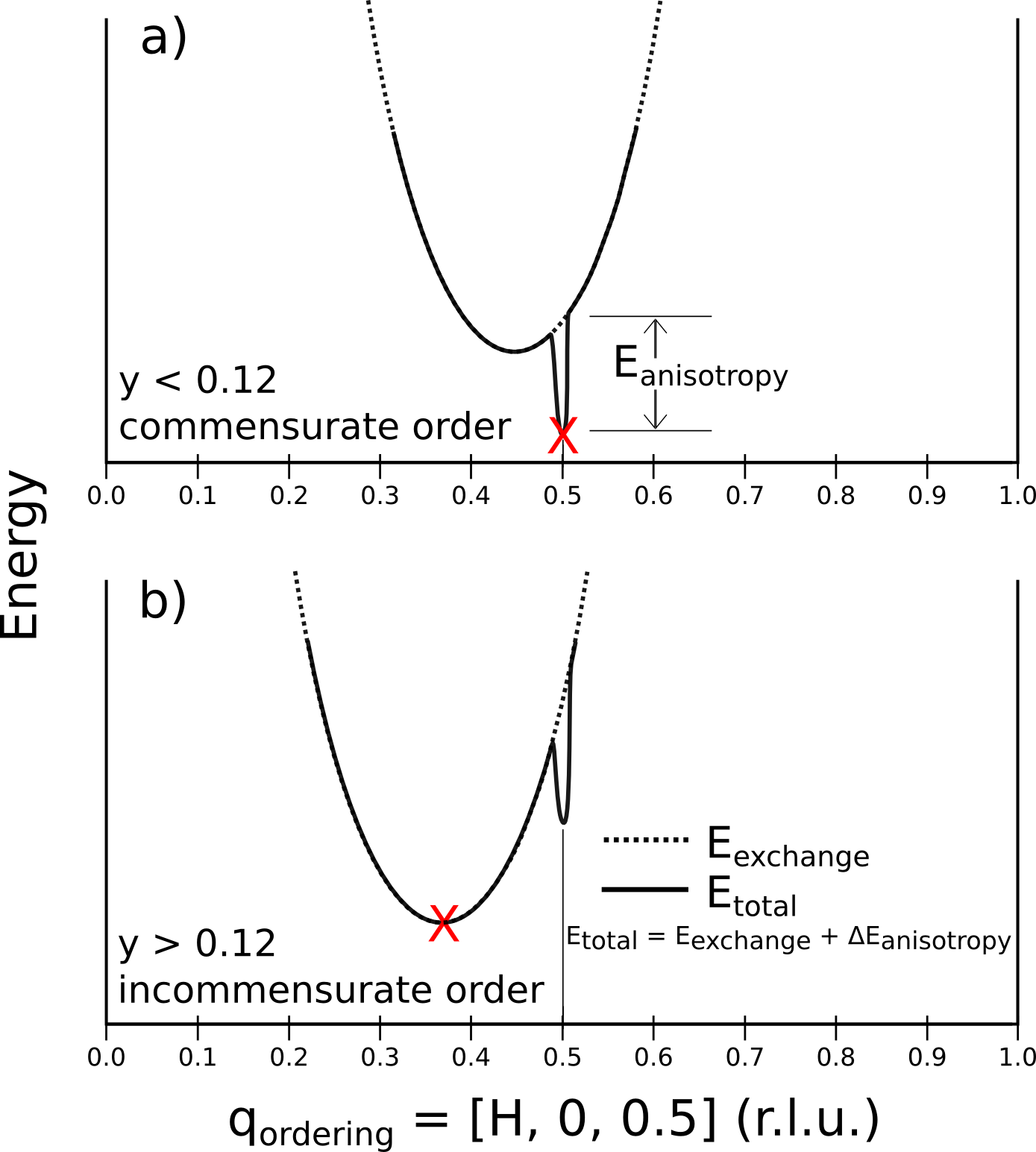} 
	\caption{ (color online)
Schematic of the  magnetic energy in \FT\ at zero temperature as a function of the ordering wavevector, for two values of the excess iron concentration $y$. The exchange energy, $E_{exchange}$, is represented as a parabola (dashed line); the structural anisotropy at the commensurate wavevector lowers the total energy, $E_{total}$, further by ${\Delta}E_{anisotropy}$.  The system always settles at the global energy minimum, marked with an ``X''.
{\bf (a)} For values of $y <$ 0.12, the minimum of $E_{exchange}$ is close to the commensurate wavevector, and so the global minimum is at the \emph{commensurate} position.  Above \TN\ the anisotropy well becomes filled, thus spin fluctuations appear at \Qinc\ = [0.45, 0, 0.5], the exchange energy minimum.
{\bf (b)} For $y >$ 0.12, the energy minimum for $E_{exchange}$ is far from the commensurate position, which puts the global minimum at an \emph{incommensurate} wavevector near \Qinc\ = [0.38, 0, $\nicefrac{1}{2}$] \cite{Bao09,StockRodriguez}.
}
   \label{fig:anisotropy}
\end{figure}

Fig. \ref{fig:anisotropy} shows a schematic of the energy as a function of the ordering wavevector, \Qorder, and illustrates the competition between the high-temperature incommensurate paramagnetic and low-temperature commensurate magnetic states. In the tetragonal phase the interactions between the electrons favor an incommensurate phase whose wavevector is not too far from [$\nicefrac{1}{2}$, 0, $\nicefrac{1}{2}$].  This is represented by a dotted parabola on the energy vs. \Qorder\ plot, which has a minimum at an incommensurate wavevector.  Here we are not concerned with the microscopic mechanism that causes the system to choose such an incommensurate wavevector (although such a wavevector is not difficult to achieve, given the interplay between different exchange, superexchange, and double-exchange pathways \cite{Turner09}).  
However, the commensurate wavevector is special, because it allows the system to lower energy by an amount ${\Delta}E_{anisotropy}$ through the monoclinic lattice distortion, which in turn allows the magnetic order to align the moments according to the spin anisotropy.  Thus $E_{total}$ has a dip at the commensurate \Qvec.  If that dip is close to the minimum of the exchange interactions, then it will be the global minimum, and the ground state will become bicollinear through the monoclininc distortion (this distortion is probably cooperative and related to the strong magnetoelastic coupling \cite{Paul2011-2}, since J$_1$ and J$_2$ will split under the monoclinic and orthorhombic distortions, respectively).  At temperatures greater than the anisotropy gap ($\approx$ 6 meV, corresponding to 69 K), the spins are thermally excited above the gap, and the advantage it provides to the commensurate order disappears.  Thus the incommensurate fluctuations we observe in the paramagnetic phase are the critical fluctuations leading up to a standard second-order transition, and in the absence of the lock-in transition the system would order at this incommensurate wavevector; but this transition is interrupted by the first-order transtition to the commensurate state.

This lock-in scenario is expected to take place in the low-Fe doping regime.  When the excess iron doping is increased beyond the critical concentration $y \approx$ 0.12, the system enters a mixed-phase regime \cite{RodriguezStock, Zaliznyak11-2}, consistent with two minima (at the commensurate and incommensurate wavevectors).  With further Fe-doping, the system becomes single-phase, and the incommensurability increases rapidly to \Qinc\ = [0.38, 0, $\nicefrac{1}{2}$] \cite{RodriguezStock}.  The orthorhombic distortion is also a function of doping, and decreases beyond the critical doping \cite{Bao09}, which suggests the reduction of anisotropy.
This can be understood as a doping-driven (as opposed to temperature-driven) classical commensurate-incommensurate transition \cite{Choi_Chen_Radzihovsky}, in which the cost of the exchange energy required to form the bicollinear phase increases with doping, eventually becoming greater than the energy gained by the anisotropy gap (see Fig. \ref{fig:anisotropy}).  Doping should change the incommensurate wavevector preferred by the exchange energy; for lower values of iron doping with $y <$ 0.12, this dependence can be found by  examining the fluctuations just above \TN\ (below \TN, the system is of course gapped and commensurate).
This doping-driven commensurate-incommensurate transition is expected to be continuous and can be interpreted as the condensation of the domain walls (or solitons) of the commensurate state.  One consequence of this transition is that at finite temperature a soliton liquid is expected to occur between the commensurate state (soliton vacuum) and the incommensurate state (soliton line crystal) 
\cite{Coppersmith1982, Pokrovsky1986}.
The recent observation of anomalous hysteresis in the thermal expansion of Fe$_{1.13}$Te has been attributed to strong soliton pinning \cite{Roessler2011}.

To conclude, we have observed clear signatures of a competition between commensurate and incommensurate magnetism in a well-characterized sample of \samp.  We find incommensurate fluctuations above \TN, which give way to commensurate order below \TN\ with a gap in the excitation spectrum. This behavior, as well as other previously unexplained observations, can be understood in terms of a lock-in transition induced by the spin anisotropy gap which is present in the monoclinic bicollinear phase and absent in the tetragonal phase.

\begin{acknowledgments}
D.P and D.R. were supported by the DOE, Office of Basic Energy Sciences under Contract No. DE-SC0006939.  The extended visit of D.P. to KIT, during which part of this work was performed, was supported by the International Institute for Complex Adaptive Matter, and by NSF grants DMR-0847782, DMR-0820579, and DMR-0844115.  L.R. and G.C. were supported through NSF grant No. DMR-1001240.  The authors are grateful to I. Mazin and S. Maekawa for valuable discussions, and to J. L. Niedziela for critical reading of the manuscript.
\end{acknowledgments}

%

\begin{thebibliography}{10}%
\makeatletter
\providecommand \@ifxundefined [1]{%
 \ifx #1\undefined \expandafter \@firstoftwo
 \else \expandafter \@secondoftwo
\fi
}%
\providecommand \@ifnum [1]{%
 \ifnum #1\expandafter \@firstoftwo
 \else \expandafter \@secondoftwo
\fi
}%
\providecommand \enquote [1]{``#1''}%
\providecommand \bibnamefont  [1]{#1}%
\providecommand \bibfnamefont [1]{#1}%
\providecommand \citenamefont [1]{#1}%
\providecommand\href[0]{\@sanitize\@href}%
\providecommand\@href[1]{\endgroup\@@startlink{#1}\endgroup\@@href}%
\providecommand\@@href[1]{#1\@@endlink}%
\providecommand \@sanitize [0]{\begingroup\catcode`\&12\catcode`\#12\relax}%
\@ifxundefined \pdfoutput {\@firstoftwo}{%
 \@ifnum{\z@=\pdfoutput}{\@firstoftwo}{\@secondoftwo}%
}{%
 \providecommand\@@startlink[1]{\leavevmode}%
 \providecommand\@@endlink[0]{}%
}{%
 \providecommand\@@startlink[1]{%
  \leavevmode
  \pdfstartlink
   attr{/Border[0 0 1 ]/H/I/C[0 1 1]}%
   user{/Subtype/Link/A<</Type/Action/S/URI/URI(#1)>>}%
  \relax
 }%
 \providecommand\@@endlink[0]{\pdfendlink}%
}%
\providecommand \url  [0]{\begingroup\@sanitize \@url }%
\providecommand \@url [1]{\endgroup\@href {#1}{\urlprefix}}%
\providecommand \urlprefix [0]{URL }%
\providecommand \Eprint[0]{\href }%
\@ifxundefined \urlstyle {%
  \providecommand \doi [1]{doi:\discretionary{}{}{}#1}%
}{%
  \providecommand \doi [0]{doi:\discretionary{}{}{}\begingroup
  \urlstyle{rm}\Url }%
}%
\providecommand \doibase [0]{http://dx.doi.org/}%
\providecommand \Doi[1]{\href{\doibase#1}}%
\providecommand \bibAnnote [3]{%
  \BibitemShut{#1}%
  \begin{quotation}\noindent
    \textsc{Key:}\ #2\\\textsc{Annotation:}\ #3%
  \end{quotation}%
}%
\providecommand \bibAnnoteFile [2]{%
  \IfFileExists{#2}{\bibAnnote {#1} {#2} {\input{#2}}}{}%
}%
\providecommand \typeout [0]{\immediate \write \m@ne }%
\providecommand \selectlanguage [0]{\@gobble}%
\providecommand \bibinfo [0]{\@secondoftwo}%
\providecommand \bibfield [0]{\@secondoftwo}%
\providecommand \translation [1]{[#1]}%
\providecommand \BibitemOpen[0]{}%
\providecommand \bibitemStop [0]{}%
\providecommand \bibitemNoStop [0]{.\EOS\space}%
\providecommand \EOS [0]{\spacefactor3000\relax}%
\providecommand \BibitemShut [1]{\csname bibitem#1\endcsname}%
\bibitem{Kamihara2008}%
  \BibitemOpen
  \bibfield{author}{%
  \bibinfo {author} {\bibfnamefont{Y.}~\bibnamefont{Kamihara}}, \bibinfo
  {author} {\bibfnamefont{T.}~\bibnamefont{Watanabe}}, \bibinfo {author}
  {\bibfnamefont{M.}~\bibnamefont{Hirano}},\ and\ \bibinfo {author}
  {\bibfnamefont{H.}~\bibnamefont{Hosono}},\ }%
  \bibfield{journal}{%
  \Doi{10.1021/ja800073m}{\bibinfo {journal} {Journal of the American Chemical
  Society}}\ }%
  \textbf{\bibinfo {volume} {130}},\ \bibinfo {pages} {3296} (\bibinfo {year}
  {2008}),\
  \Eprint{http://arxiv.org/abs/http://pubs.acs.org/doi/pdf/10.1021/ja800073m}{%
http://pubs.acs.org/doi/pdf/10.1021/ja800073m},\
  \url{http://pubs.acs.org/doi/abs/10.1021/ja800073m}%
  \bibAnnoteFile{NoStop}{Kamihara2008}%
\bibitem{Paglione2010}%
  \BibitemOpen
  \bibfield{author}{%
  \bibinfo {author} {\bibfnamefont{J.}~\bibnamefont{Paglione}}\ and\ \bibinfo
  {author} {\bibfnamefont{R.~L.}\ \bibnamefont{Greene}},\ }%
  \bibfield{journal}{%
  \Doi{10.1038/nphys1759}{\bibinfo {journal} {Nature Physics}}\ }%
  \textbf{\bibinfo {volume} {6}},\ \bibinfo {pages} {645} (\bibinfo {year}
  {2010})%
  \bibAnnoteFile{NoStop}{Paglione2010}%
\bibitem{Mazin08}%
  \BibitemOpen
  \bibfield{author}{%
  \bibinfo {author} {\bibfnamefont{I.~I.}\ \bibnamefont{Mazin}}, \bibinfo
  {author} {\bibfnamefont{D.~J.}\ \bibnamefont{Singh}}, \bibinfo {author}
  {\bibfnamefont{M.~D.}\ \bibnamefont{Johannes}},\ and\ \bibinfo {author}
  {\bibfnamefont{M.~H.}\ \bibnamefont{Du}},\ }%
  \bibfield{journal}{%
  \Doi{10.1103/PhysRevLett.101.057003}{\bibinfo {journal} {Physical Review
  Letters}}\ }%
  \textbf{\bibinfo {volume} {101}},\ \bibinfo {pages} {057003} (\bibinfo {year}
  {2008})%
  \bibAnnoteFile{NoStop}{Mazin08}%
\bibitem{Singh08}%
  \BibitemOpen
  \bibfield{author}{%
  \bibinfo {author} {\bibfnamefont{D.~J.}\ \bibnamefont{Singh}}\ and\ \bibinfo
  {author} {\bibfnamefont{M.-H.}\ \bibnamefont{Du}},\ }%
  \bibfield{journal}{%
  \Doi{10.1103/PhysRevLett.100.237003}{\bibinfo {journal} {Physical Review
  Letters}}\ }%
  \textbf{\bibinfo {volume} {100}},\ \bibinfo {pages} {237003} (\bibinfo {year}
  {2008})%
  \bibAnnoteFile{NoStop}{Singh08}%
\bibitem{Bao09}%
  \BibitemOpen
  \bibfield{author}{%
  \bibinfo {author} {\bibfnamefont{W.}~\bibnamefont{Bao}}, \bibinfo {author}
  {\bibfnamefont{Y.}~\bibnamefont{Qiu}}, \bibinfo {author}
  {\bibfnamefont{Q.}~\bibnamefont{Huang}}, \bibinfo {author}
  {\bibfnamefont{M.~A.}\ \bibnamefont{Green}}, \bibinfo {author}
  {\bibfnamefont{P.}~\bibnamefont{Zajdel}}, \bibinfo {author}
  {\bibfnamefont{M.~R.}\ \bibnamefont{Fitzsimmons}}, \bibinfo {author}
  {\bibfnamefont{M.}~\bibnamefont{Zhernenkov}}, \bibinfo {author}
  {\bibfnamefont{S.}~\bibnamefont{Chang}}, \bibinfo {author}
  {\bibfnamefont{M.}~\bibnamefont{Fang}}, \bibinfo {author}
  {\bibfnamefont{B.}~\bibnamefont{Qian}}, \bibinfo {author}
  {\bibfnamefont{E.~K.}\ \bibnamefont{Vehstedt}}, \bibinfo {author}
  {\bibfnamefont{J.}~\bibnamefont{Yang}}, \bibinfo {author}
  {\bibfnamefont{H.~M.}\ \bibnamefont{Pham}}, \bibinfo {author}
  {\bibfnamefont{L.}~\bibnamefont{Spinu}},\ and\ \bibinfo {author}
  {\bibfnamefont{Z.~Q.}\ \bibnamefont{Mao}},\ }%
  \bibfield{journal}{%
  \Doi{10.1103/PhysRevLett.102.247001}{\bibinfo {journal} {Physical Review
  Letters}}\ }%
  \textbf{\bibinfo {volume} {102}},\ \bibinfo {pages} {247001} (\bibinfo {year}
  {2009})%
  \bibAnnoteFile{NoStop}{Bao09}%
\bibitem{Subedi08}%
  \BibitemOpen
  \bibfield{author}{%
  \bibinfo {author} {\bibfnamefont{A.}~\bibnamefont{Subedi}}, \bibinfo {author}
  {\bibfnamefont{L.}~\bibnamefont{Zhang}}, \bibinfo {author}
  {\bibfnamefont{D.}~\bibnamefont{Singh}},\ and\ \bibinfo {author}
  {\bibfnamefont{M.}~\bibnamefont{Du}},\ }%
  \bibfield{journal}{%
  \Doi{10.1103/PhysRevB.78.134514}{\bibinfo {journal} {Physical Review B}}\ }%
  \textbf{\bibinfo {volume} {78}},\ \bibinfo {pages} {134514} (\bibinfo {year}
  {2008})%
  \bibAnnoteFile{NoStop}{Subedi08}%
\bibitem{Xia2009}%
  \BibitemOpen
  \bibfield{author}{%
  \bibinfo {author} {\bibfnamefont{Y.}~\bibnamefont{Xia}}, \bibinfo {author}
  {\bibfnamefont{D.}~\bibnamefont{Qian}}, \bibinfo {author}
  {\bibfnamefont{L.}~\bibnamefont{Wray}}, \bibinfo {author}
  {\bibfnamefont{D.}~\bibnamefont{Hsieh}}, \bibinfo {author}
  {\bibfnamefont{G.~F.}\ \bibnamefont{Chen}}, \bibinfo {author}
  {\bibfnamefont{J.~L.}\ \bibnamefont{Luo}}, \bibinfo {author}
  {\bibfnamefont{N.~L.}\ \bibnamefont{Wang}},\ and\ \bibinfo {author}
  {\bibfnamefont{M.~Z.}\ \bibnamefont{Hasan}},\ }%
  \bibfield{journal}{%
  \Doi{10.1103/PhysRevLett.103.037002}{\bibinfo {journal} {Phys. Rev. Lett.}}\
  }%
  \textbf{\bibinfo {volume} {103}},\ \bibinfo {pages} {037002} (\bibinfo
  {month} {Jul}\ \bibinfo {year} {2009})%
  \bibAnnoteFile{NoStop}{Xia2009}%
\bibitem{Turner09}%
  \BibitemOpen
  \bibfield{author}{%
  \bibinfo {author} {\bibfnamefont{A.~M.}\ \bibnamefont{Turner}}, \bibinfo
  {author} {\bibfnamefont{F.}~\bibnamefont{Wang}},\ and\ \bibinfo {author}
  {\bibfnamefont{A.}~\bibnamefont{Vishwanath}},\ }%
  \bibfield{journal}{%
  \Doi{10.1103/PhysRevB.80.224504}{\bibinfo {journal} {Physical Review B}}\ }%
  \textbf{\bibinfo {volume} {80}},\ \bibinfo {pages} {224504} (\bibinfo {year}
  {2009})%
  \bibAnnoteFile{NoStop}{Turner09}%
\bibitem{Liu10}%
  \BibitemOpen
  \bibfield{author}{%
  \bibinfo {author} {\bibfnamefont{T.~J.}\ \bibnamefont{Liu}}, \bibinfo
  {author} {\bibfnamefont{J.}~\bibnamefont{Hu}}, \bibinfo {author}
  {\bibfnamefont{B.}~\bibnamefont{Qian}}, \bibinfo {author}
  {\bibfnamefont{D.}~\bibnamefont{Fobes}}, \bibinfo {author}
  {\bibfnamefont{Z.~Q.}\ \bibnamefont{Mao}}, \bibinfo {author}
  {\bibfnamefont{W.}~\bibnamefont{Bao}}, \bibinfo {author}
  {\bibfnamefont{M.}~\bibnamefont{Reehuis}}, \bibinfo {author}
  {\bibfnamefont{S.~A.~J.}\ \bibnamefont{Kimber}}, \bibinfo {author}
  {\bibfnamefont{K.}~\bibnamefont{Proke{\v{s}}}}, \bibinfo {author}
  {\bibfnamefont{S.}~\bibnamefont{Matas}}, \bibinfo {author}
  {\bibfnamefont{D.~N.}\ \bibnamefont{Argyriou}}, \bibinfo {author}
  {\bibfnamefont{A.}~\bibnamefont{Hiess}}, \bibinfo {author}
  {\bibfnamefont{A.}~\bibnamefont{Rotaru}}, \bibinfo {author}
  {\bibfnamefont{H.}~\bibnamefont{Pham}}, \bibinfo {author}
  {\bibfnamefont{L.}~\bibnamefont{Spinu}}, \bibinfo {author}
  {\bibfnamefont{Y.}~\bibnamefont{Qiu}}, \bibinfo {author}
  {\bibfnamefont{V.}~\bibnamefont{Thampy}}, \bibinfo {author}
  {\bibfnamefont{A.~T.}\ \bibnamefont{Savici}}, \bibinfo {author}
  {\bibfnamefont{J.~A.}\ \bibnamefont{Rodriguez}},\ and\ \bibinfo {author}
  {\bibfnamefont{C.}~\bibnamefont{Broholm}},\ }%
  \bibfield{journal}{%
  \Doi{10.1038/nmat2800}{\bibinfo {journal} {Nature Materials}}\ }%
  \textbf{\bibinfo {volume} {9}},\ \bibinfo {pages} {718} (\bibinfo {year}
  {2010})%
  \bibAnnoteFile{NoStop}{Liu10}%
\bibitem{Lipscombe11}%
  \BibitemOpen
  \bibfield{author}{%
  \bibinfo {author} {\bibfnamefont{O.}~\bibnamefont{Lipscombe}}, \bibinfo
  {author} {\bibfnamefont{G.}~\bibnamefont{Chen}}, \bibinfo {author}
  {\bibfnamefont{C.}~\bibnamefont{Fang}}, \bibinfo {author}
  {\bibfnamefont{T.}~\bibnamefont{Perring}}, \bibinfo {author}
  {\bibfnamefont{D.}~\bibnamefont{Abernathy}}, \bibinfo {author}
  {\bibfnamefont{A.}~\bibnamefont{Christianson}}, \bibinfo {author}
  {\bibfnamefont{T.}~\bibnamefont{Egami}}, \bibinfo {author}
  {\bibfnamefont{N.}~\bibnamefont{Wang}}, \bibinfo {author}
  {\bibfnamefont{J.}~\bibnamefont{Hu}},\ and\ \bibinfo {author}
  {\bibfnamefont{P.}~\bibnamefont{Dai}},\ }%
  \bibfield{journal}{%
  \Doi{10.1103/PhysRevLett.106.057004}{\bibinfo {journal} {Physical Review
  Letters}}\ }%
  \textbf{\bibinfo {volume} {106}},\ \bibinfo {pages} {057004} (\bibinfo {year}
  {2011})%
  \bibAnnoteFile{NoStop}{Lipscombe11}%
\bibitem{StockRodriguez}%
  \BibitemOpen
  \bibfield{author}{%
  \bibinfo {author} {\bibfnamefont{C.}~\bibnamefont{Stock}}, \bibinfo {author}
  {\bibfnamefont{E.~E.}\ \bibnamefont{Rodriguez}}, \bibinfo {author}
  {\bibfnamefont{P.}~\bibnamefont{Zavalij}}, \bibinfo {author}
  {\bibfnamefont{M.~A.}\ \bibnamefont{Green}},\ and\ \bibinfo {author}
  {\bibfnamefont{J.~A.}\ \bibnamefont{Rodriguez-Rivera}},\ }%
  \bibfield{journal}{%
  \bibinfo {journal} {arXiv:1103.1811}}%
   (\bibinfo {year} {2011}),\
  \Eprint{http://arxiv.org/abs/1103.1811).}{1103.1811).}%
  \bibAnnoteFile{Stop}{StockRodriguez}%
\bibitem{Li09-1}%
  \BibitemOpen
  \bibfield{author}{%
  \bibinfo {author} {\bibfnamefont{S.}~\bibnamefont{Li}}, \bibinfo {author}
  {\bibfnamefont{C.}~\bibnamefont{de~la Cruz}}, \bibinfo {author}
  {\bibfnamefont{Q.}~\bibnamefont{Huang}}, \bibinfo {author}
  {\bibfnamefont{Y.}~\bibnamefont{Chen}}, \bibinfo {author}
  {\bibfnamefont{J.~W.}\ \bibnamefont{Lynn}}, \bibinfo {author}
  {\bibfnamefont{J.}~\bibnamefont{Hu}}, \bibinfo {author}
  {\bibfnamefont{Y.-L.}\ \bibnamefont{Huang}}, \bibinfo {author}
  {\bibfnamefont{F.-C.}\ \bibnamefont{Hsu}}, \bibinfo {author}
  {\bibfnamefont{K.-W.}\ \bibnamefont{Yeh}}, \bibinfo {author}
  {\bibfnamefont{M.-K.}\ \bibnamefont{Wu}},\ and\ \bibinfo {author}
  {\bibfnamefont{P.}~\bibnamefont{Dai}},\ }%
  \bibfield{journal}{%
  \Doi{10.1103/PhysRevB.79.054503}{\bibinfo {journal} {Physical Review B}}\ }%
  \textbf{\bibinfo {volume} {79}},\ \bibinfo {pages} {054503} (\bibinfo {year}
  {2009})%
  \bibAnnoteFile{NoStop}{Li09-1}%
\bibitem{Wojdyr10}%
  \BibitemOpen
  \bibfield{author}{%
  \bibinfo {author} {\bibfnamefont{M.}~\bibnamefont{Wojdyr}},\ }%
  \bibfield{journal}{%
  \Doi{10.1107/S0021889810030499}{\bibinfo {journal} {Journal of Applied
  Crystallography}}\ }%
  \textbf{\bibinfo {volume} {43}},\ \bibinfo {pages} {1126} (\bibinfo {year}
  {2010})%
  \bibAnnoteFile{NoStop}{Wojdyr10}%
\bibitem{Ma09}%
  \BibitemOpen
  \bibfield{author}{%
  \bibinfo {author} {\bibfnamefont{F.}~\bibnamefont{Ma}}, \bibinfo {author}
  {\bibfnamefont{W.}~\bibnamefont{Ji}}, \bibinfo {author}
  {\bibfnamefont{J.}~\bibnamefont{Hu}}, \bibinfo {author}
  {\bibfnamefont{Z.-Y.}\ \bibnamefont{Lu}},\ and\ \bibinfo {author}
  {\bibfnamefont{T.}~\bibnamefont{Xiang}},\ }%
  \bibfield{journal}{%
  \Doi{10.1103/PhysRevLett.102.177003}{\bibinfo {journal} {Physical Review
  Letters}}\ }%
  \textbf{\bibinfo {volume} {102}},\ \bibinfo {pages} {177003} (\bibinfo {year}
  {2009})%
  \bibAnnoteFile{NoStop}{Ma09}%
\bibitem{Choi_Chen_Radzihovsky}%
  \BibitemOpen
  \bibfield{author}{%
  \bibinfo {author} {\bibfnamefont{S.}~\bibnamefont{Choi}}, \bibinfo {author}
  {\bibfnamefont{G.}~\bibnamefont{Chen}},\ and\ \bibinfo {author}
  {\bibfnamefont{L.}~\bibnamefont{Radzihovsky}},\ }%
  \bibinfo {journal} {unpublished}%
  \bibAnnoteFile{NoStop}{Choi_Chen_Radzihovsky}%
\bibitem{Lipscombe10}%
  \BibitemOpen
\bibfield{journal}{%
    }%
  \bibfield{author}{%
  \bibinfo {author} {\bibfnamefont{O.}~\bibnamefont{Lipscombe}}, \bibinfo
  {author} {\bibfnamefont{L.}~\bibnamefont{Harriger}}, \bibinfo {author}
  {\bibfnamefont{P.}~\bibnamefont{Freeman}}, \bibinfo {author}
  {\bibfnamefont{M.}~\bibnamefont{Enderle}}, \bibinfo {author}
  {\bibfnamefont{C.}~\bibnamefont{Zhang}}, \bibinfo {author}
  {\bibfnamefont{M.}~\bibnamefont{Wang}}, \bibinfo {author}
  {\bibfnamefont{T.}~\bibnamefont{Egami}}, \bibinfo {author}
  {\bibfnamefont{J.}~\bibnamefont{Hu}}, \bibinfo {author}
  {\bibfnamefont{T.}~\bibnamefont{Xiang}}, \bibinfo {author}
  {\bibfnamefont{M.}~\bibnamefont{Norman}},\ and\ \bibinfo {author}
  {\bibfnamefont{P.}~\bibnamefont{Dai}},\ }%
  \bibfield{journal}{%
  \bibinfo {journal} {Physical Review B}\ }%
  \textbf{\bibinfo {volume} {82}} (\bibinfo {year} {2010}),\ \doi{\bibinfo
  {doi} {10.1103/PhysRevB.82.064515}}%
  \bibAnnoteFile{NoStop}{Lipscombe10}%
\bibitem{Paul2011-2}%
  \BibitemOpen
  \bibfield{author}{%
  \bibinfo {author} {\bibfnamefont{I.}~\bibnamefont{Paul}}, \bibinfo {author}
  {\bibfnamefont{A.}~\bibnamefont{Cano}},\ and\ \bibinfo {author}
  {\bibfnamefont{K.}~\bibnamefont{Sengupta}},\ }%
  \bibfield{journal}{%
  \Doi{10.1103/PhysRevB.83.115109}{\bibinfo {journal} {Phys. Rev. B}}\ }%
  \textbf{\bibinfo {volume} {83}},\ \bibinfo {pages} {115109} (\bibinfo {month}
  {Mar}\ \bibinfo {year} {2011}),\
  \url{http://link.aps.org/doi/10.1103/PhysRevB.83.115109}%
  \bibAnnoteFile{NoStop}{Paul2011-2}%
\bibitem{RodriguezStock}%
  \BibitemOpen
  \bibfield{author}{%
  \bibinfo {author} {\bibfnamefont{E.~E.}\ \bibnamefont{Rodriguez}}, \bibinfo
  {author} {\bibfnamefont{C.}~\bibnamefont{Stock}}, \bibinfo {author}
  {\bibfnamefont{P.}~\bibnamefont{Zajdel}}, \bibinfo {author}
  {\bibfnamefont{K.~L.}\ \bibnamefont{Krycka}}, \bibinfo {author}
  {\bibfnamefont{C.~F.}\ \bibnamefont{Majkrzak}}, \bibinfo {author}
  {\bibfnamefont{P.}~\bibnamefont{Zavalij}},\ and\ \bibinfo {author}
  {\bibfnamefont{M.~A.}\ \bibnamefont{Green}},\ }%
  \bibfield{journal}{%
  \bibinfo {journal} {Physical Review B}\ }%
  \textbf{\bibinfo {volume} {84}},\ \bibinfo {pages} {064403} (\bibinfo {year}
  {2011}),\ \Eprint{http://arxiv.org/abs/1012.0590v1}{1012.0590v1}%
  \bibAnnoteFile{NoStop}{RodriguezStock}%
\bibitem{Zaliznyak11-2}%
  \BibitemOpen
  \bibfield{author}{%
  \bibinfo {author} {\bibfnamefont{I.~A.}\ \bibnamefont{Zaliznyak}}, \bibinfo
  {author} {\bibfnamefont{Z.~J.}\ \bibnamefont{Xu}}, \bibinfo {author}
  {\bibfnamefont{J.~S.}\ \bibnamefont{Wen}}, \bibinfo {author}
  {\bibfnamefont{J.~M.}\ \bibnamefont{Tranquada}}, \bibinfo {author}
  {\bibfnamefont{G.~D.}\ \bibnamefont{Gu}}, \bibinfo {author}
  {\bibfnamefont{V.}~\bibnamefont{Solovyov}}, \bibinfo {author}
  {\bibfnamefont{V.~N.}\ \bibnamefont{Glazkov}}, \bibinfo {author}
  {\bibfnamefont{A.~I.}\ \bibnamefont{Zheludev}}, \bibinfo {author}
  {\bibfnamefont{V.~O.}\ \bibnamefont{Garlea}},\ and\ \bibinfo {author}
  {\bibfnamefont{M.~B.}\ \bibnamefont{Stone}},\ }%
  \bibfield{journal}{%
  \bibinfo {journal} {arXiv:1108.5968v1}}%
   (\bibinfo {year} {2012})%
  \bibAnnoteFile{NoStop}{Zaliznyak11-2}%
\bibitem{Coppersmith1982}%
  \BibitemOpen
  \bibfield{author}{%
  \bibinfo {author} {\bibfnamefont{S.~N.}\ \bibnamefont{Coppersmith}}, \bibinfo
  {author} {\bibfnamefont{D.~S.}\ \bibnamefont{Fisher}}, \bibinfo {author}
  {\bibfnamefont{B.~I.}\ \bibnamefont{Halperin}}, \bibinfo {author}
  {\bibfnamefont{P.~A.}\ \bibnamefont{Lee}},\ and\ \bibinfo {author}
  {\bibfnamefont{W.~F.}\ \bibnamefont{Brinkman}},\ }%
  \bibfield{journal}{%
  \Doi{10.1103/PhysRevB.25.349}{\bibinfo {journal} {Physical Review B}}\ }%
  \textbf{\bibinfo {volume} {25}},\ \bibinfo {pages} {349} (\bibinfo {month}
  {Jan}\ \bibinfo {year} {1982}),\
  \url{http://link.aps.org/doi/10.1103/PhysRevB.25.349}%
  \bibAnnoteFile{NoStop}{Coppersmith1982}%
\bibitem{Pokrovsky1986}%
  \BibitemOpen
  \bibfield{author}{%
  \bibinfo {author} {\bibfnamefont{V.}~\bibnamefont{Pokrovsky}}\ and\ \bibinfo
  {author} {\bibnamefont{et~al.}},\ }%
  \enquote{\bibinfo {title} {Solitons},}\ \ (\bibinfo {publisher} {North
  Holland, Amsterdam},\ \bibinfo {year} {1986})\ pp.\ \bibinfo {pages}
  {71--127}%
  \bibAnnoteFile{NoStop}{Pokrovsky1986}%
\bibitem{Roessler2011}%
  \BibitemOpen
  \bibfield{author}{%
  \bibinfo {author} {\bibfnamefont{S.}~\bibnamefont{R\"o\ss{}ler}}, \bibinfo
  {author} {\bibfnamefont{D.}~\bibnamefont{Cherian}}, \bibinfo {author}
  {\bibfnamefont{W.}~\bibnamefont{Lorenz}}, \bibinfo {author}
  {\bibfnamefont{M.}~\bibnamefont{Doerr}}, \bibinfo {author}
  {\bibfnamefont{C.}~\bibnamefont{Koz}}, \bibinfo {author}
  {\bibfnamefont{C.}~\bibnamefont{Curfs}}, \bibinfo {author}
  {\bibfnamefont{Y.}~\bibnamefont{Prots}}, \bibinfo {author}
  {\bibfnamefont{U.~K.}\ \bibnamefont{R\"o\ss{}ler}}, \bibinfo {author}
  {\bibfnamefont{U.}~\bibnamefont{Schwarz}}, \bibinfo {author}
  {\bibfnamefont{S.}~\bibnamefont{Elizabeth}},\ and\ \bibinfo {author}
  {\bibfnamefont{S.}~\bibnamefont{Wirth}},\ }%
  \bibfield{journal}{%
  \Doi{10.1103/PhysRevB.84.174506}{\bibinfo {journal} {Phys. Rev. B}}\ }%
  \textbf{\bibinfo {volume} {84}},\ \bibinfo {pages} {174506} (\bibinfo {month}
  {Nov}\ \bibinfo {year} {2011}),\
  \url{http://link.aps.org/doi/10.1103/PhysRevB.84.174506}%
  \bibAnnoteFile{NoStop}{Roessler2011}%
\end{thebibliography}
\end{document}